\documentclass[showpacs,preprintnumbers,amsmath,amssymb,floatfix]{revtex4-1}

\usepackage{graphicx}

\usepackage{dcolumn}

\usepackage{bm}

\begin{document}

\title{Einstein and M\o ller energy-momentum complexes for a new regular
black hole solution with a nonlinear electrodynamics source}

\author{I. Radinschi}
\email{radinschi@yahoo.com } \affiliation{Department of Physics ,
 Gh. Asachi   Technical University,
Iasi, 700050, Romania}

\author{Farook Rahaman}
\email{rahaman@associates.iucaa.in } \affiliation{Department of
Mathematics, Jadavpur University, Kolkata 700 032, West Bengal,
India}

\author{Th. Grammenos}
\email{thgramme@uth.gr } \affiliation{Department of Civil Engineering,
University of Thessaly, 383 34 Volos, Greece}

\author{  Sayeedul Islam }
\email{  sayeedul.jumath@gmail.com } \affiliation{Department of
Mathematics, Jadavpur University, Kolkata 700 032, West Bengal,
India.}

\date{\today}

\begin{abstract}
A study about the energy and momentum distributions of a new charged regular black hole
solution with a nonlinear electrodynamics source is presented. The energy
and momentum are calculated using the Einstein and M\o ller energy-momentum
complexes. The results show that in both pseudotensorial prescriptions the
expressions for the energy of the gravitational background depend on the
mass $M$ and the charge $q$ of the black hole, an additional factor $\beta $
coming from the spacetime metric considered, and the radial coordinate $r$,
while in both prescriptions all the momenta vanish. Further, it is pointed
out that in some limiting and particular cases the two complexes yield the
same expression for the energy distribution as that obtained in the relevant
literature for the Schwarzschild black hole solution.
\end{abstract}

\maketitle

\section{Introduction}

Energy-momentum localization plays a leading role in the theories advanced
over the years in relation to General Relativity. There is a major
difficulty, however, in formulating a proper definition for the energy
density of gravitational backgrounds. Indeed, the key problem is the lack of
a satisfactory description for the gravitational energy.

Many researchers have conducted extensive research using different methods
for energy-momentum localization. Standard research methods include the use
of different tools, such as super-energy tensors [1], quasi-local
expressions [2] and the famous energy-momentum complexes of Einstein [3],
Landau-Lifshitz [4], Papapetrou [5], Bergmann-Thomson [6], M\o ller [7],
Weinberg [8], and Qadir-Sharif [9]. The main problem encountered is the dependence on the reference frame of these pseudotensorial
prescriptions. An alternative method used in many studies on computing the
energy and momentum distributions in order to avoid the dependence on
coordinates is the teleparallel theory of gravitation [10].

As regards pseudotensorial prescriptions, only the M\o ller energy-momentum
complex is a coordinate independent tool. Schwarzschild Cartesian
coordinates and Kerr-Schild Cartesian coordinates are useful to compute the
energy-momentum in the case of the other pseudotensorial definitions. Over
the past few decades, despite the criticism directed against energy-momentum
complexes concerning mainly the physicalness of the results obtained by them, their application has provided physically reasonable results for
many spacetime geometries, more particularly for geometries in ($3+1$), ($%
2+1 $) and ($1+1$) dimensions [11]-[12].

There is an agreement between the Einstein, Landau-Lifshitz, Papapetrou,
Bergmann-Thomson, Weinberg and M\o ller prescriptions, on the one hand, and
the definition of the quasi-local mass advanced by Penrose [13] and
developed by Tod [14] for some gravitating systems, on the other hand (see [15] for a comprehensive review).
Several pseudotensorial definitions ``provide the same results'' for any
metric of the Kerr-Schild class and for solutions that are more general than
the Kerr-Schild class (see, for example, the works of J. M. Aguirregabiria,
A. Chamorro and K. S. Virbhadra, and S. S. Xulu in [11], and K. S. Virbhadra
in [16]). Furthermore, the similarity between some of the aforementioned
results and those obtained by using the teleparallel theory of gravitation [17]
  cannot be overlooked. In fact, the history of energy-momentum complexes
should include their definition and use, as well as the attempts for their rehabilitation [18].

The present work has the following structure: in Section 2 we describe the new
spherically symmetric, static, charged regular black hole solution with a
nonlinear electrodynamics source [19] under study. Section 3 is focused on
the presentation of the Einstein and M\o ller energy-momentum complexes used
for performing the calculations. Section 4 contains the calculations of the
energy and momentum distributions. In the Discussion and Final Remarks given
in Section 5, we make a brief description of the results of our
investigation as well as some limiting and particular cases. Throughout the
article we use geometrized units ($c=G=1$), the signature chosen for our
purpose is ($+$,$-$,$-$,$-$), and the calculations are performed using the
Schwarzschild Cartesian coordinates $\{t$, $x$, $y$, $z\}$ for the Einstein
prescription and the Schwarzschild coordinates $\{t,$ $r,$ $\theta ,$ $%
\varphi \}$ for the M\o ller prescription. Also, Greek indices range from $0$
to $3$, while Latin indices run from $1$ to $3$.

\section{Description of the New Regular Black Hole Solution with a Nonlinear
Electrodynamics Source}

In this section we present a new spherically symmetric, static, charged regular
black hole solution with a nonlinear electrodynamics source recently
developed by L. Balart and E. C. Vagenas [19]  and
we analyze the distribution of energy and momentum using the
Einstein and M\o ller prescriptions.

A brief but interesting discussion about the regular black hole solutions
that have been obtained by coupling gravity to nonlinear electrodynamics
theories is presented in [19] (see Introduction and References therein for
more details).  Further, in an interesting and similar work the horizon entropy of a black hole
is determined as a function of Komar energy and the horizon
area [20].

In order to develop the new charged regular black hole solution, the authors of
[19] considered the Fermi-Dirac-type distribution. For this purpose, they
generalized the methodology developed in Sec. 2 of their paper by
considering distribution functions raised to the power of a real number
greater than zero. We notice that the methodology presented in Sec. 2
consists in constructing a general charged regular black hole metric for mass
distribution functions that are inspired by continuous probability density
distributions. The corresponding electric field for each black hole solution
is also constructed in terms of a general mass distribution function. The
metric function is given by
\begin{equation}
f(r)=1-\frac{2\,M}{r}\left(\frac{\sigma (\beta \,r)}{\sigma _{\infty }}\right)^{\beta },
\tag{1}
\end{equation}
where $\sigma _{\infty }= \sigma(r\rightarrow \infty )$ is a normalization
factor and the function $\sigma (\beta \,r)$ corresponds to any one of the
mass functions listed in Table 1 of [19], but with the coordinate $r$
multiplied by an additional factor $\beta>0$.

The new spherically symmetric, static, charged regular black hole solution
with a nonlinear electrodynamics source given by eq. (29) in [19] is
obtained, as we pointed out above, using the Fermi-Dirac-type distribution,
and the metric function becomes now
\begin{equation}
f(r)=1-\frac{2\,M}{r}\left(\frac{2}{\exp \left(\frac{q^{2}}{\beta \,M\,r}\right)+1}\right)^{\beta
}.  \tag{2}
\end{equation}
Moreover, when $r\rightarrow \infty $ the mass function $m(r)=\,M(\frac{%
\sigma (\beta \,r)}{\sigma _{\infty }})^{\beta }\rightarrow M$. The
distribution function satisfies the condition $\frac{\sigma (\,r)}{\sigma
_{\infty }}\rightarrow 1$ when $r\rightarrow \infty $. This solution is a
generalization of the Ay\'{o}n-Beato and Garc\'{\i}a black hole solution
[21].
\vspace{.35cm}

The corresponding electric field has the expression
\begin{equation*}
E(r)=\frac{q}{r^2}\exp\left( \frac{(1-\beta )q^2}{2\beta Mr}\right) \left[ \text{sech} \left( \frac{q^2}{2\beta Mr}\right) \right] ^{1+\beta }
\end{equation*}
\begin{equation}
\times \left[ 1-\frac{q^{2}}{4Mr}\tanh \left( \frac{q^{2}}{2\beta Mr}\right)
+\frac{1}{4\beta Mr}\left( \frac{1-\beta }{\exp\left\{ \frac{q^{2}}{\beta Mr}%
\right\} +1}\right) \right] .  \tag{3}
\end{equation}
In order to construct the extremal regular black hole metric
for this example, some values of $\beta $ and the corresponding charges are
listed in Table 2 of [19].

Finally, the new charged regular black hole solution with a nonlinear
electrodynamics source is described by the metric
\begin{equation}
ds^{2}=B(r)dt^{2}-A(r)dr^{2}-r^{2}(d\theta ^{2}+\sin^2 \theta d\phi ^{2}),
\tag{4}
\end{equation}
with $B(r)=f(r)$, $A(r)=\frac{1}{f(r)}$.
\begin{figure*}[thbp]
 \includegraphics[width=5.0cm]{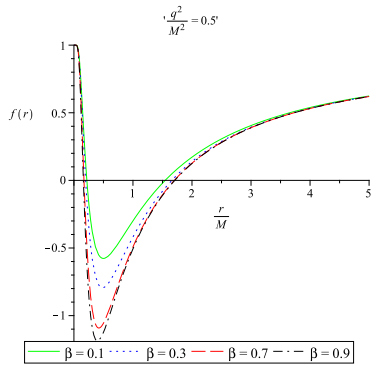}
   \centering
 \caption{
 An inner and an outer horizon exist at the points where $f(r)$ meets the $\frac{r}{M}$-axis, here shown for four different values of the parameter $\beta$.   }
\end{figure*}

Figure 1 shows that two horizons exist at the points where $f(r)$ meets the $\frac{r}{M}$-axis and for four different values of the parameter $\beta$. We have chosen $\left(\frac{q}{M}\right)^2 = 0.5$.  Note that the  positions of the inner
  and the outer horizon  remain unaffected for various values of $\beta$.

\section{Einstein and \ M\o ller Energy-Momentum Complexes}

The Einstein energy-momentum complex [3] for a ($3+1$) dimensional
gravitational background has the well-known expression
\begin{equation}
\theta _{\nu }^{\mu }=\frac{1}{16\pi }h_{\nu ,\,\lambda }^{\mu \lambda }.
\tag{5}
\end{equation}%
The superpotentials $h_{\nu }^{\mu \lambda }$ involved in (5) are given by
\begin{equation}
h_{\nu }^{\mu \lambda }=\frac{1}{\sqrt{-g}}g_{\nu \sigma }\left[-g(g^{\mu \sigma
}g^{\lambda \kappa }-g^{\lambda \sigma }g^{\mu \kappa })\right]_{,\kappa }  \tag{6}
\end{equation}%
which satisfies the necessary antisymmetric property:
\begin{equation}
h_{\nu }^{\mu \lambda }=-h_{\nu }^{\lambda \mu }.  \tag{7}
\end{equation}
In the Einstein prescription the local conservation law is respected
\begin{equation}
\theta _{\nu ,\,\mu }^{\mu }=0.  \tag{8}
\end{equation}%
Thus, the energy and momentum can be evaluated in Einstein's prescription
with
\begin{equation}
P_{\mu }=\int \int \int \theta _{\mu }^{0}\,dx^{1}dx^{2}dx^{3}.  \tag{9}
\end{equation}%
Here, $\theta _{0}^{0}$ and $\theta _{i}^{0}$ represent the energy and
momentum density components, respectively.

Applying Gauss' theorem the energy-momentum reads

\begin{equation}
P_{\mu }=\frac{1}{16\pi }\int \int  h_{\mu }^{0i}n_{i}dS,  \tag{10}
\end{equation}

with $n_{i}$ the outward unit normal vector over the surface $dS.$ In eq.
(10) $P_{0}$ is the energy.

Concerning the expression for the M{\o }ller energy-momentum complex [7] we
have

\begin{equation}
\mathcal{J}_{\nu }^{\mu }=\frac{1}{8\pi }M_{\nu \,\,,\,\lambda }^{\mu
\lambda },  \tag{11}
\end{equation}%
with the M{\o }ller superpotentials $M_{\nu }^{\mu \lambda }$ given by
\begin{equation}
M_{\nu }^{\mu \lambda }=\sqrt{-g}\left( \frac{\partial g_{\nu \sigma }}{%
\partial x^{\kappa }}-\frac{\partial g_{\nu \kappa }}{\partial x^{\sigma }}%
\right) g^{\mu \kappa }g^{\lambda \sigma }.  \tag{12}
\end{equation}%
The M{\o }ller superpotentials $M_{\nu }^{\mu \lambda }$ are antisymmetric:
\begin{equation}
M_{\nu }^{\mu \lambda }=-M_{\nu }^{\lambda \mu }.  \tag{13}
\end{equation}%
Like the Einstein energy-momentum complex, M{\o }ller's energy-momentum
complex also satisfies the local conservation law%
\begin{equation}
\frac{\partial \mathcal{J}_{\nu }^{\mu }}{\partial x^{\mu }}=0.  \tag{14}
\end{equation}%
In (14) $\mathcal{J}_{0}^{0}$ gives the energy density and $\mathcal{J}%
_{i}^{0}$ represents the momentum density components.

For the M\o ller prescription, the energy and momentum distributions are
obtained by
\begin{equation}
P_{\mu }=\int \int \int \mathcal{J}_{\mu }^{0}dx^{1}dx^{2}dx^{3}  \tag{15}
\end{equation}%
and the energy distribution can be calculated by
\begin{equation}
E=\int \int \int \mathcal{J}_{0}^{0}dx^{1}dx^{2}dx^{3}.  \tag{16}
\end{equation}%
\qquad Again, using Gauss' theorem one gets

\begin{equation}
P_{\mu }=\frac{1}{8\pi }\int \int  M_{\mu }^{0i}n_{i}dS.  \tag{17}
\end{equation}

\section{Energy and Momentum Distribution for the New Regular Black Hole
Solution with a Nonlinear Electrodynamics Source}

In order to compute the energy and momenta in the Einstein prescription, it
is useful to transform the metric given by the line element (4) in
Schwarzschild Cartesian coordinates applying the coordinate transformation $%
x=r\,\sin \theta \cos \varphi ,$ $y=r\,\sin \theta \sin \varphi ,$ $%
z=r\,\cos \theta $. Then, the following line element is obtained:

\begin{equation}
ds^{2}=B(r)dt^{2}-(dx^{2}+dy^{2}+dz^{2})-\frac{A(r)-1}{r^{2}}%
(xdx+ydy+zdz)^{2}\text{.}  \tag{18}
\end{equation}

The components of the superpotential $h_{\mu }^{0i}$ in quasi-Cartesian
coordinates for $\mu =1,2,3$ and $i=1,2,3$ are given by

\[
h_{1}^{01}  = h_{1}^{02}=h_{1}^{03}=0,  \]
\[~~~~~
~~~~~~~~~~~~~~~ ~~~~~~~~~~~~~~~~~~~~~~~~~~~~~~~~ ~~~~~ ~~~~~ h_{2}^{01}  = h_{2}^{02}=h_{2}^{03}=0,~ ~~~~~~~~~~ ~~~~~~~~~~~~~~~~~~~~~~~~~~~~~~~~~~~~~~~~~~~
  ~~ ~(19)\]
\[h_{3}^{01}  = h_{3}^{02}=h_{3}^{03}=0.  \]

Now, using (6) we compute the non-vanishing components of the
superpotentials in the Einstein prescription and we obtain the following
expressions:

\begin{equation}
h_{0}^{01}=\frac{2\,x}{r^{2}}\frac{2\,M}{r}\left(\frac{2}{\exp \left(\frac{q^{2}}{%
\beta \,M\,r}\right)+1}\right)^{\beta },  \tag{20}
\end{equation}

\begin{equation}
h_{0}^{02}=\frac{2\,y}{r^{2}}\frac{2\,M}{r}\left(\frac{2}{\exp \left(\frac{q^{2}}{%
\beta \,M\,r}\right)+1}\right)^{\beta },  \tag{21}
\end{equation}

\begin{equation}
h_{0}^{03}=\frac{2\,z}{r^{2}}\frac{2\,M}{r}\left(\frac{2}{\exp \left(\frac{q^{2}}{%
\beta \,M\,r}\right)+1}\right)^{\beta }.  \tag{22}
\end{equation}

Combining the line element (18), the expression for the energy from (10) and the expressions
(20)-(22) for the superpotentials, one obtains the energy distribution for
the new charged regular  black hole in the Einstein prescription:

\begin{equation}
E_{E}=M\left(\frac{2}{\exp \left(\frac{q^{2}}{\beta \,M\,r}\right)+1}\right)^{\beta }.  \tag{23}
\end{equation}

In order to get the momentum components we use (10) and (19) and performing
the calculations we find that all the momenta are zero:

\begin{equation}
P_{x}=P_{y}=P_{z}=0.  \tag{24}
\end{equation}
In the left panel of  Fig. 2,   we plot the energy distribution in the Einstein prescription for
different values of $\beta $ and  $\left(\frac{q}{M}\right)^2 = 0.5$.

Using the M\o ller prescription, which is applied in Schwarzschild
coordinates $\{t,$ $r,$ $\theta ,$ $\varphi \}$, the only non-vanishing
superpotential is given by

\begin{equation}
M_{0}^{01}=\left(\frac{2\,M\left(\frac{2}{\exp \left(\frac{q^{2}}{\beta M\,r}\right)+1}\right)^{\beta }%
}{r^{2}}-\frac{2\,\left(\frac{2}{\exp \left(\frac{q^{2}}{\beta \,M\,r}\right)+1}\right)^{\beta
}\times \exp \left(\frac{q^{2}}{\beta \,M\,r}\right)\times q^{2}}{r^{3}\left(\exp\left (\frac{%
q^{2}}{\beta \,M\,r}\right)+1\right)}\right)r^{2}\,\sin \theta ,  \tag{25}
\end{equation}
while all the other components of the M\o ller superpotential vanish.

Applying the aforementioned result for the line element (4) and using the
expression (17) for the energy, we calculate the energy distribution in the M%
\o ller prescription:

\begin{equation}
E_{M}=\left[\frac{2}{\exp \left(\frac{q^{2}}{\beta \,M\,r}\right)+1%
}\right]^{\beta}\left[M-\frac{q^{2}\exp\left(\frac{q^{2}}{\beta\,M\,r}\right)}{r\left(\exp\left (\frac{%
q^{2}}{\beta \,M\,r}\right)+1\right)}\right] . \tag{26}
\end{equation}
Our calculations yield for all the momenta:

\begin{equation}
P_{r}=P_{\theta }=P_{\varphi }=0.  \tag{27}
\end{equation}

In the right panel of  Fig. 2 we plot the energy distribution in the M\o ller prescription for
different values of $\beta $ and $\left(\frac{q}{M}\right)^2 = 0.5$. Now, in Fig. 3,  we compare  the expressions for energy in the Einstein and M\o %
ller prescriptions for $\beta =0.7$. Note that for large radial distances the values of energy in both prescriptions coincide.

\begin{figure*}[thbp]
\begin{tabular}{rl}
\includegraphics[width=4.0cm]{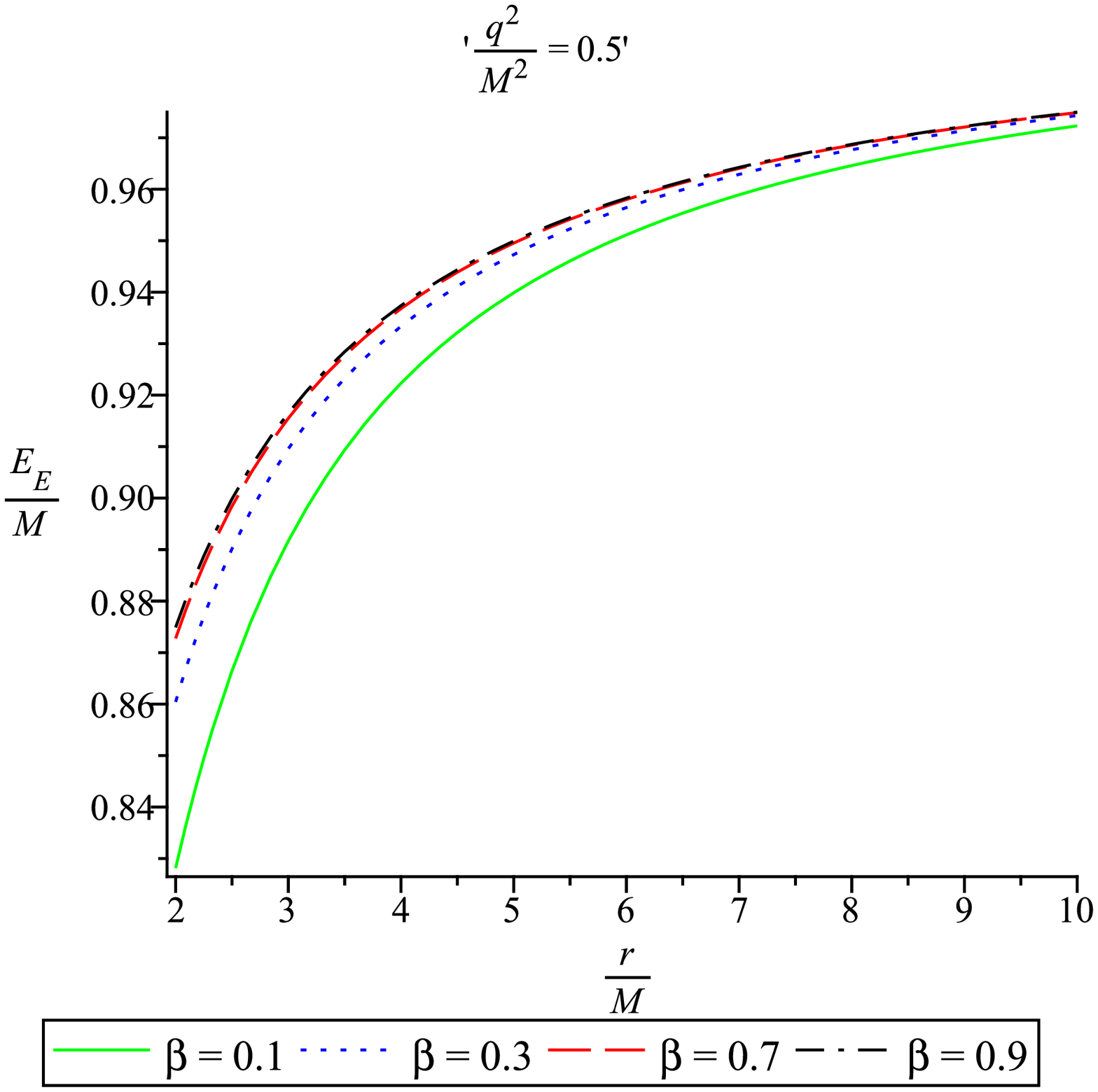}&
\includegraphics[width=4.0cm]{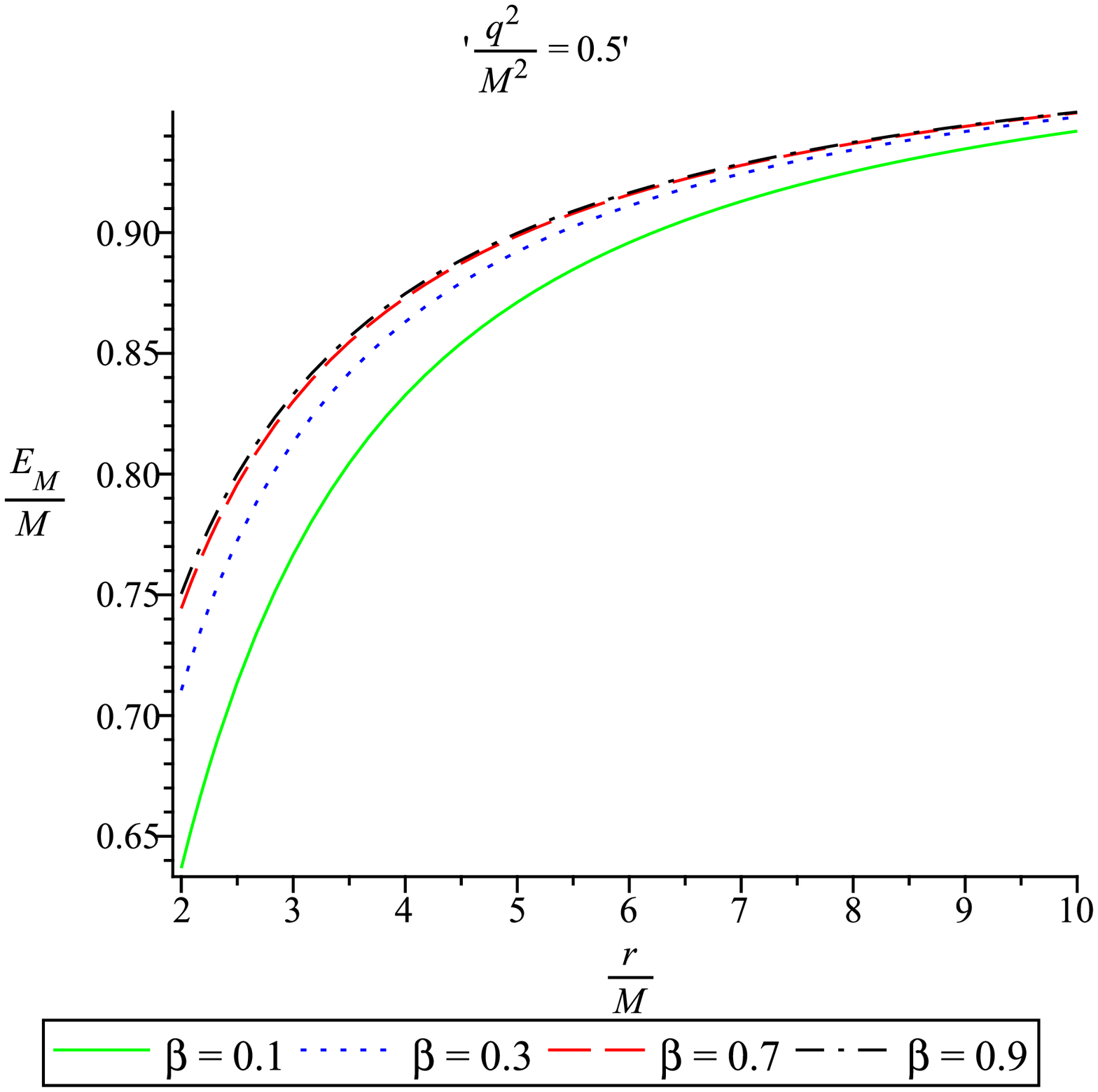}\\
\end{tabular}
\caption{ (Left) Energy distribution computed by the Einstein prescription
outside the outer horizon for four different values of $\beta$.    (Right)  Energy distribution computed by the M\o ller  prescription
outside the outer horizon for four different values of $\beta$.}
\end{figure*}
\begin{figure*}[thbp]
\begin{tabular}{rl}
\includegraphics[width=4.0cm]{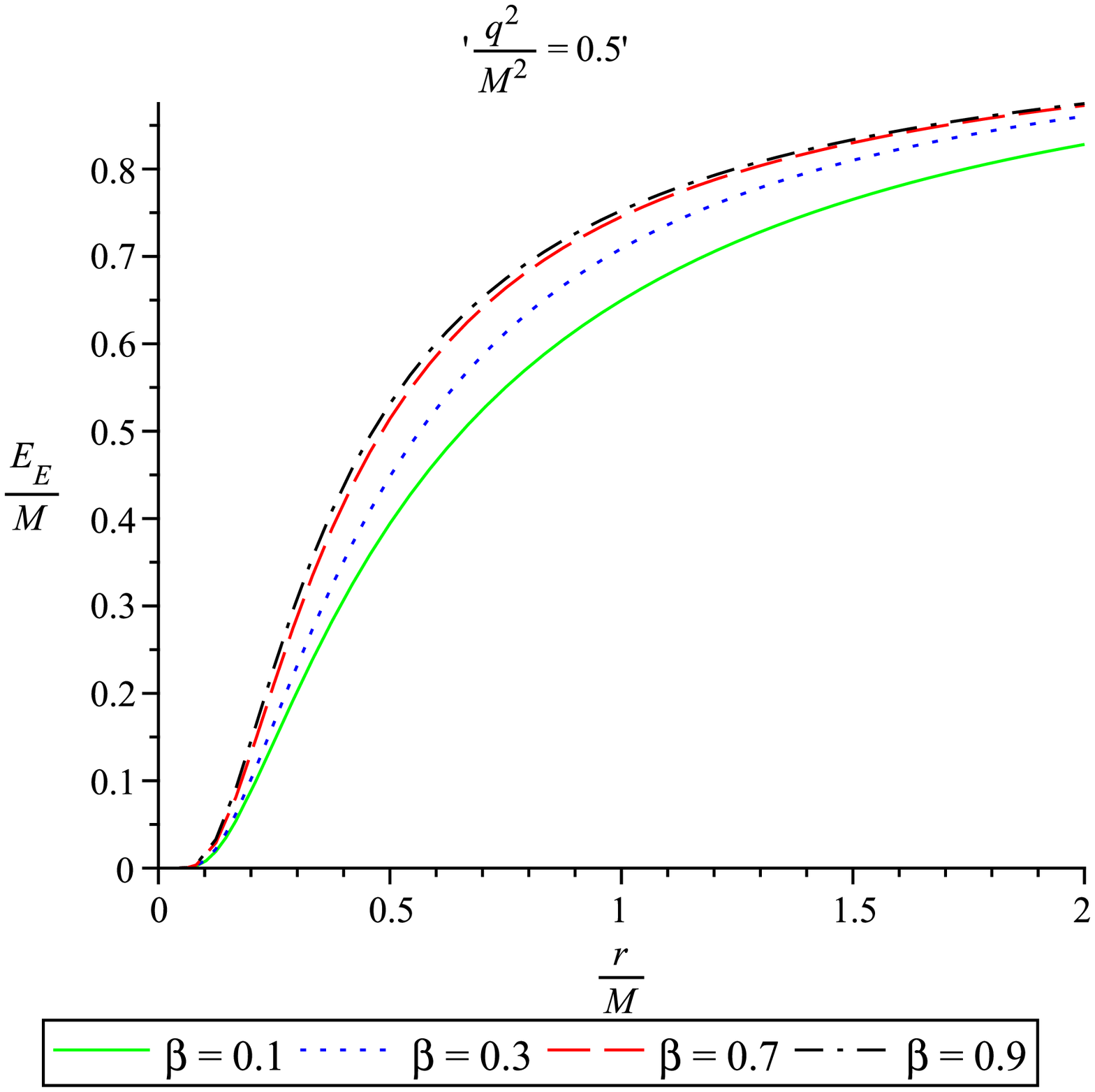}&
\includegraphics[width=4.0cm]{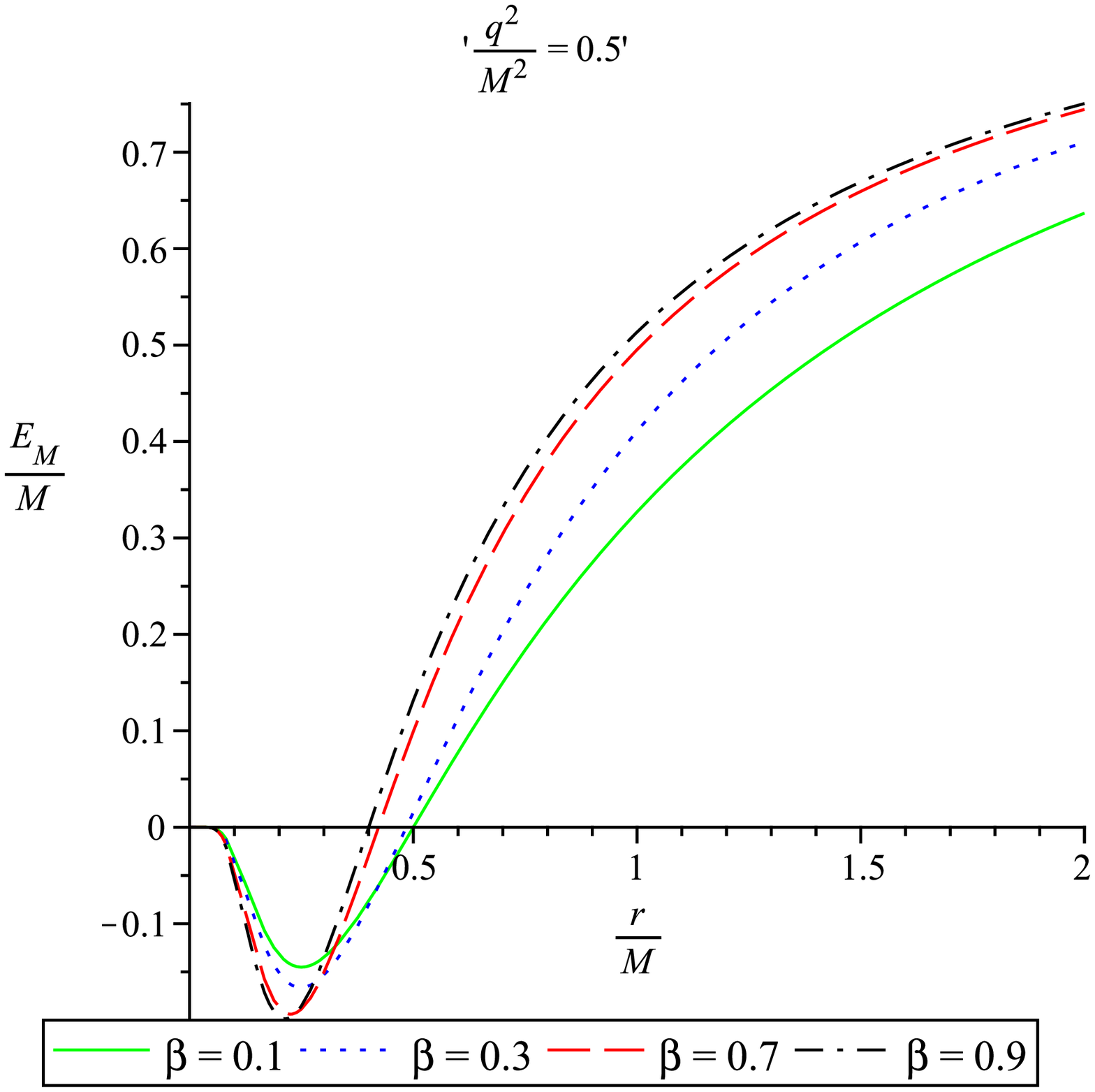}\\
\end{tabular}
\caption{   Energy distribution computed by the Einstein prescription (left) and the M\o ller prescription (right) near the origin for different values of   $\beta$ and $\left(\frac{q}{M}\right)^2 = 0.5$. }
\end{figure*}

\begin{figure*}[thbp]
\begin{tabular}{rl}
\includegraphics[width=4.5cm]{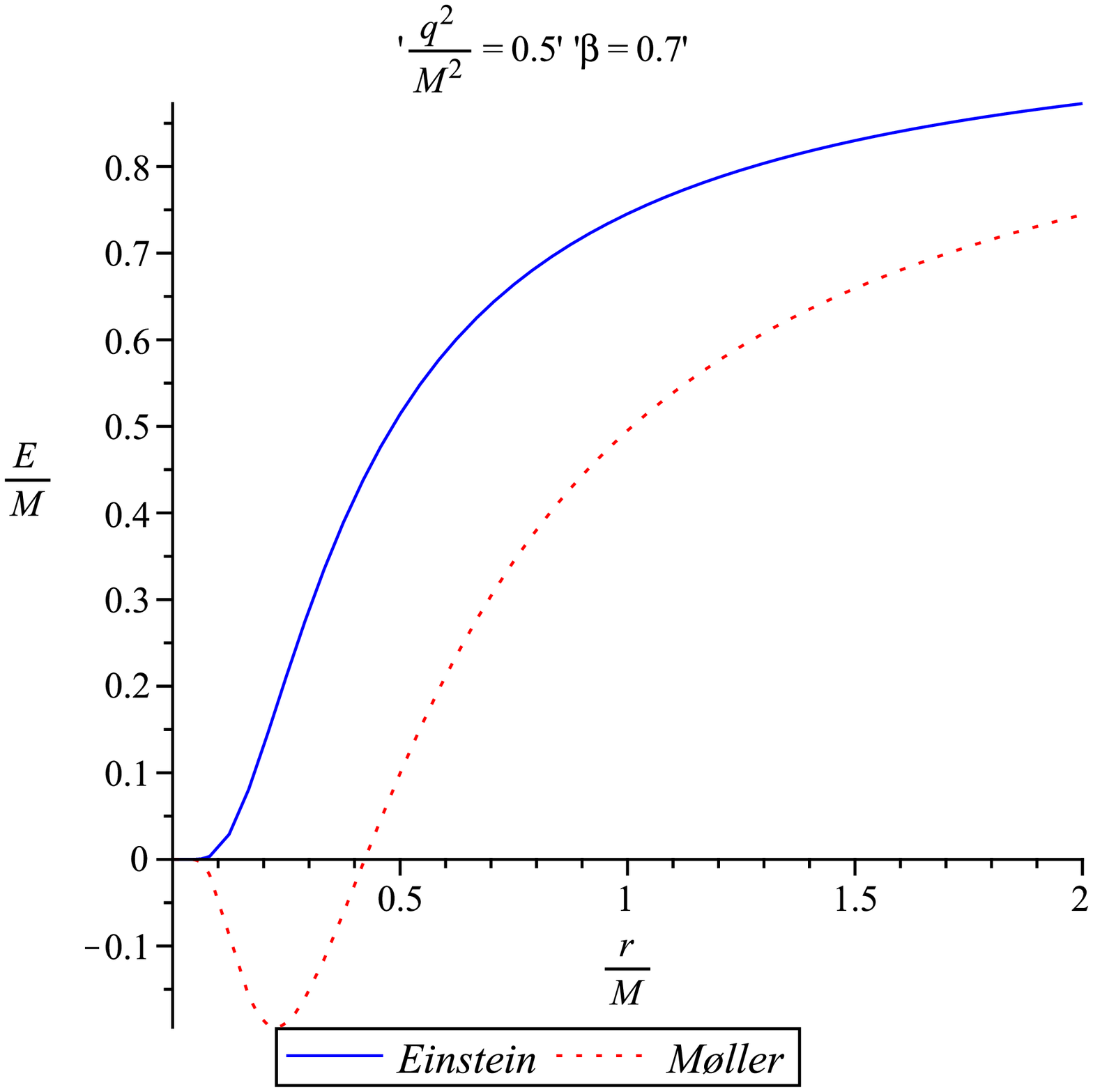}&
\includegraphics[width=4.5cm]{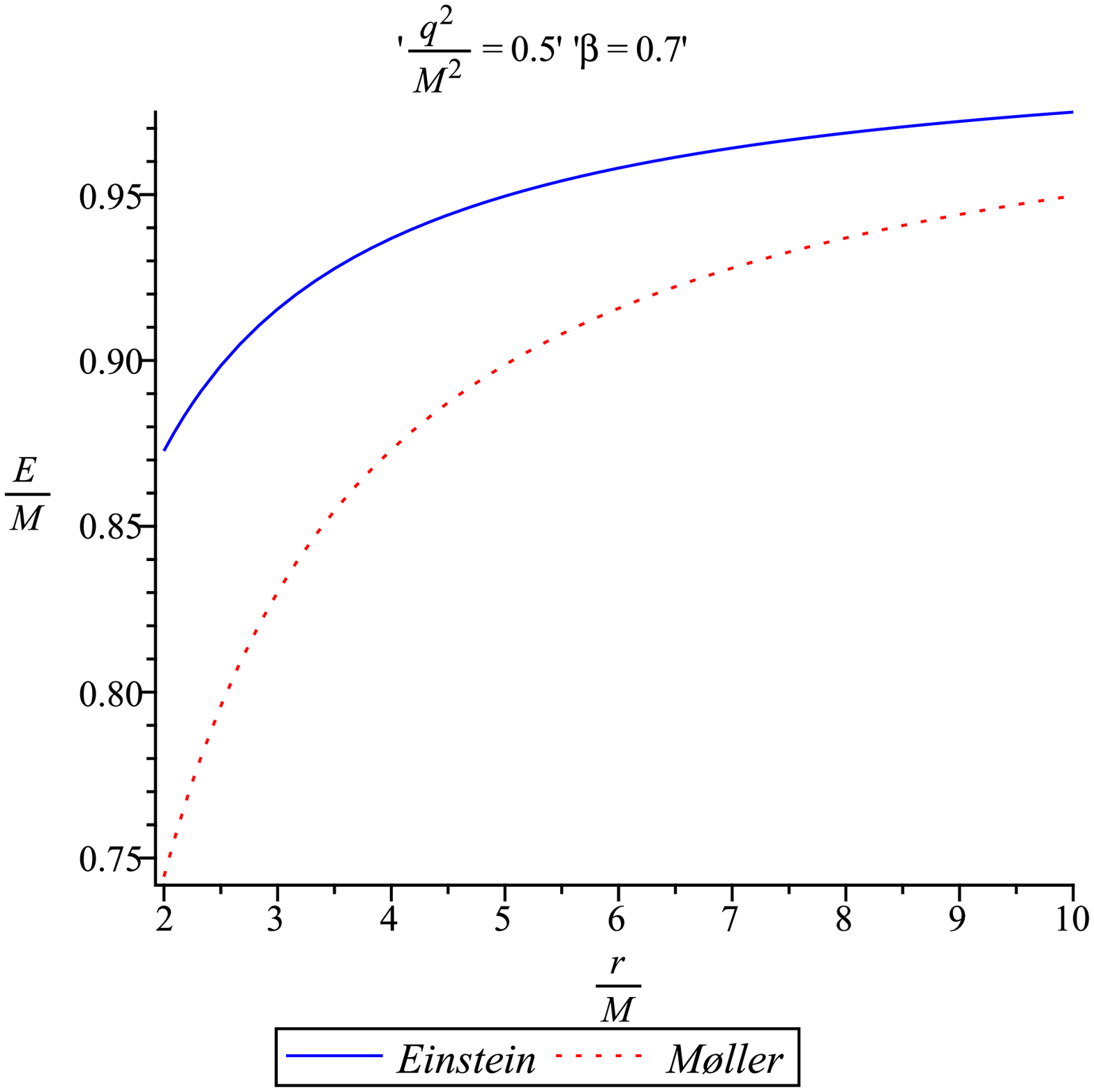}
\includegraphics[width=4.5cm]{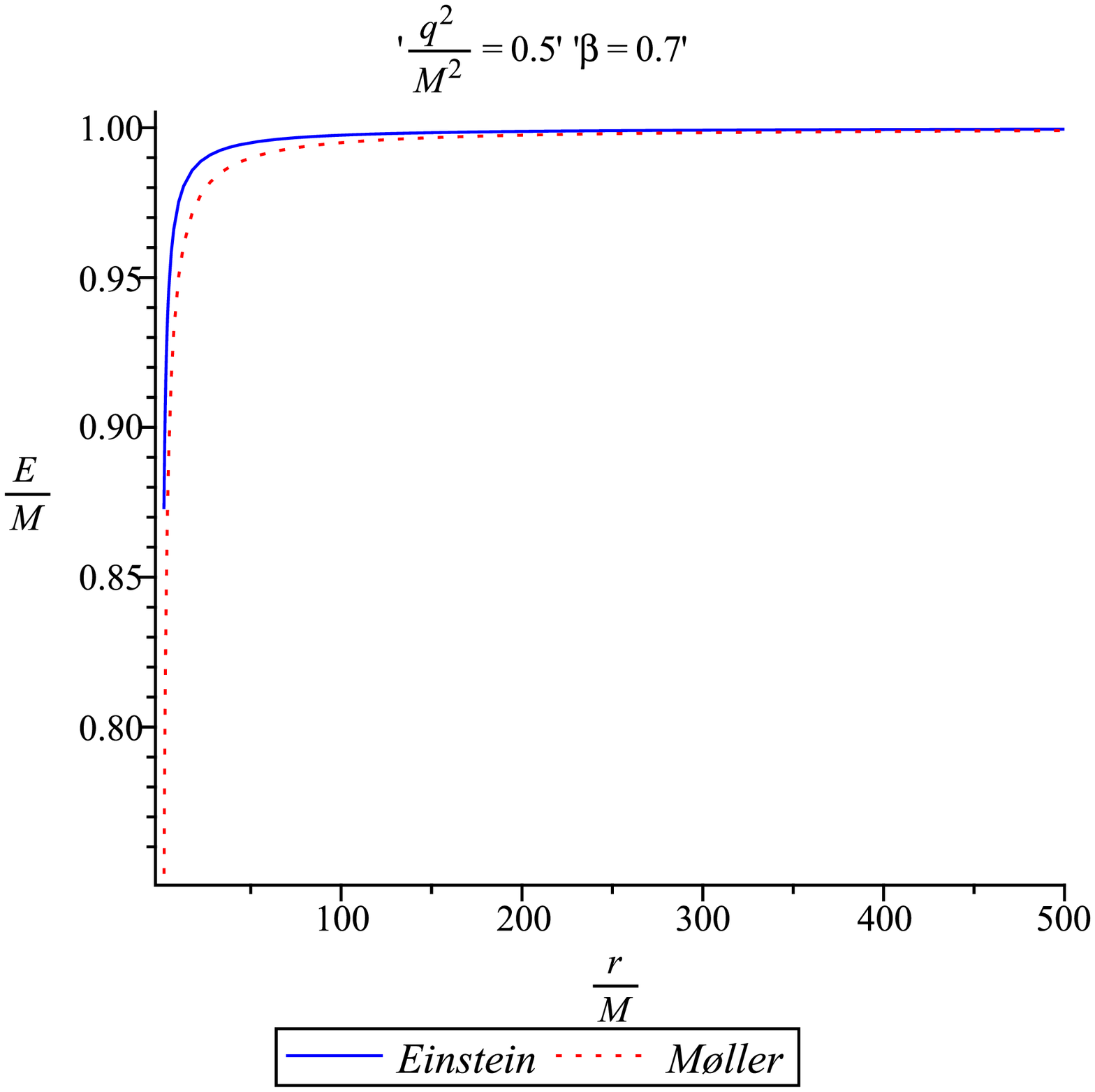}\\
\end{tabular}
\caption{ For the fixed value of $\beta=0.7$ and $\left(\frac{q}{M}\right)^2 = 0.5$: (Left) Comparison of Einstein and M\o ller energies near the origin. (Middle) Comparison of the energy distribution performed by the Einstein and M\o ller
 prescriptions outside the outer horizon.    (Right)  Comparison of the energy distribution performed by the Einstein
and M\o ller prescriptions for very large values of $r$.}
\end{figure*}
\section{Discussion and Final Remarks}
The purpose of our paper is to study the energy-momentum for a new
spherically symmetric, static and charged, regular black hole solution with a
nonlinear electrodynamics source using the Einstein and M\o ller
energy-momentum complexes. From the calculations we conclude that in the
Einstein and M\o ller prescriptions one obtains well-defined expressions
for the energy that depend on the mass $M$ of the black hole, its charge $q$,
the additional factor $\beta $ and on the radial coordinate $r$. The
calculations yield that for both the aforesaid used pseudotensorial
prescriptions all the momenta vanish.

Concerning the physical meaning of the energy-momentum expressions of
Einstein and M\o ller we study the limiting behavior of the energy for $%
r\rightarrow \infty $, $\beta \rightarrow 0$ and $\beta \rightarrow \infty $%
, and for the particular case $q=0$. The physically meaningful results for
these limiting and particular cases are presented in the following Table:
\begin{equation*}
\begin{tabular}{|c|c|c|c|c|}
\hline
Case & $r\rightarrow \infty $ & $q=0$ & $\beta \rightarrow 0$ & $\beta
\rightarrow \infty $ \\
\hline
Einstein & $M$ & $M$ & $M\exp (-\frac{q^{2}}{M\,r})$ & $M\exp \left(-\frac{q^{2}}{%
2\,M\,r}\right)$ \\
M\o ller & $M$ & $M$ & $[M-\frac{q^2}{r}]\exp \left(-\frac{q^{2}}{M\,r}\right)$
&
$[M-\frac{q^2}{2r}]\exp \left(-\frac{q^{2}}{2M\,r}\right)$\\
\hline
\end{tabular}%
\end{equation*}
\begin{equation*}
\text{Table1}
\end{equation*}

Now, some remarks are in order. Making a comparison of the results obtained
for the energy distribution with the applied Einstein and M\o ller
definitions, we conclude that for $q=0$ and at infinity $r\rightarrow \infty
$ these definitions give the same result (the ADM mass $M$) as that obtained
for the Schwarzschild black hole solution. Moreover, this is a confirmation
of Virbhadra's view point [16].

In the limiting cases $\beta \rightarrow 0$ and $\beta \rightarrow \infty $
the Einstein and M\o ller prescriptions provide different results. It is
worth noticing that there is a difference in a factor of 2 between the
exponents in the expressions for the energy. Also, we point out that
although the results are different, the expressions for the energy\ depend
on the same parameters $M$, $q$, $\beta $ and $r$. Moreover, for $\beta
\rightarrow \infty $ in both prescriptions the expressions for
energy are obtained  for the case of eq. (17) in [19]. The metric given by eq. (17) in [19]
represents a new well-known black hole solution in the literature that
contains the metric function $f(r)=1-\frac{2\,M}{r}\exp (-\frac{q^{2}}{%
2\,M\,r})$. Interestingly, if in this case we consider $r\rightarrow \infty  $  or
 $q=0$ we obtain in both prescriptions of Einstein and M\o ller the same
 expression for energy which is equal to the ADM mass $M$. In the
limiting case  $\beta \rightarrow 0$ with the metric function given by  $%
f(r)=1-\frac{2\,M}{r}\exp (-\frac{q^{2}}{M\,r})$ we deduce, after some
calculations, that the same results $E_{\text{Einstein}}=M$ and $E_{\text{M%
\o ller}}=M$ are also obtained by considering $r\rightarrow \infty $
or $q=0$. Table 2 summarizes these\ results, whereby we denote $E_{\text{%
Einstein}}$ by $E_{\text{E}}$ and $E_{\text{M\o ller}}$ by $E_{\text{M}}.$

\begin{equation*}
\begin{tabular}{|c|c|c|c|c|}
\hline
Case & $r\rightarrow \infty $   & $q=0$ & $r\rightarrow \infty $  & $\ q=0$
\\
\hline
$f(r)=1-\frac{2\,M}{r}\exp (-\frac{q^{2}}{2\,M\,r})$ & $E_{\text{E}}=M$ & $%
E_{\text{E}}=M$ & $E_{\text{M}}=M$ & $E_{\text{M}}=M$ \\
$f(r)=1-\frac{2\,M}{r}\exp (-\frac{q^{2}}{M\,r})$ & $E_{\text{E}}=M$ & $E_{%
\text{E}}=M$ & $E_{\text{M}}=M$ & $E_{\text{M}}=M$\\
\hline
\end{tabular}%
\end{equation*}

\begin{equation*}
\text{Table 2}
\end{equation*}

A final remark regarding the behavior of energy as   $r\rightarrow 0$ is deemed necessary. As one can see in Fig. 3, the energy obtained by the Einstein prescription tends to zero, while the energy obtained by the M\o ller prescription exhibits a rather strange behavior as it takes negative values in the interval $0<r<0.5$  for different values of $\beta$.

In the left panel of  Fig. 4, the comparison of the two energies, here presented for a specific value of $\beta$, shows that the energies satisfy the inequality $E_E>E_M$ as the radial distance  grows, while they tend to become equal outside the horizon for very large values of  $r$ (see Fig. 4, right panel).

The negativity of the M\o ller energy near the origin and, in fact, inside the inner horizon seems to be pathological and it could be attributed to the sensitivity of the M\o ller energy-momentum complex to the nonlinear character of the electrodynamics source. In contrast, the Einstein energy seems to be more ``shielded'' against this nonlinearity. We notice that a similar negativity behavior of the M\o ller energy has been found in [22].

In the light of the aforementioned results for the energy distribution it is
obvious that the Einstein and M\o ller energy momentum complexes provide
well-defined and physically meaningful results and are
reliable prescriptions which can be used for the study of the energy momentum
localization of gravitational backgrounds.

One can also ask what kind of astrophysical implications our results could
have. It would be possible to investigate whether the effective
gravitational mass is positive or negative by identifying the energy at
radial distance $r$ [23] with the effective gravitational mass of the
astrophysical object considered inside the region determined by the
distance $r$.    But it does not
seem that the present case is accessible to astrophysical observations since
the negative mass region is "inside the inner horizon." So the  present case is interesting  for positive effective gravitational mass of the
astrophysical object.
 Moreover, we would decide whether the astrophysical object
could act as a convergent or as a divergent gravitational lens [24-25].

Encouraged by these results, we plan, as a future perspective, to calculate
the energy-momentum of this new charged regular black hole solution by using other
energy-momentum complexes as well as the tele-parallel equivalent. These
studies can further contribute to the ongoing debate on the problem of the
energy-momentum localization.

\section*{Acknowledgments}

FR is grateful to the Inter-University Centre for Astronomy
and Astrophysics (IUCAA), India for providing Associateship
Programme. FR and SI are thankful to DST, Govt. of India for providing financial support
under SERB and INSPIRE programme. We are thankful to the referee for his constructive suggestions.


\begin{thebibliography}{99}
\bibitem{1} L. Bel, C. R. Acad. Sci. Paris \textbf{246}, 3105 (1958); I.
Robinson, Class. Quantum Grav. \textbf{14}, A331 (1997); M.A.G. Bonilla and
J.M.M. Senovilla, Gen. Rel. Grav. \textbf{29}, 91 (1997); J.M.M. Senovilla,
Class. Quantum Grav. \textbf{17}, 2799 (2000).

\bibitem{2} J. D. Brown and J.W. York, Phys. Rev. \textbf{D47}, 1407 (1993);
Sean A. Hayward, Phys. Rev. \textbf{D49}, 831 (1994); C. M. Chen, J. M.
Nester, Class. Quantum Grav. \textbf{16}, 1279 (1999), C-C.M. Liu and S. T.
Yau, Phys. Rev. Lett. \textbf{90}, 231102 (2003); L. Balart, Phys. Lett.
\textbf{B687}, 280 (2010).

\bibitem{3} A. Einstein, Preuss. Akad. Wiss. Berlin \textbf{47}, 778 (1915);
Addendum-ibid. \textbf{47}, 799 (1915); A. Trautman, in \textit{Gravitation: an
Introduction to Current Research}, ed. L. Witten (Wiley, New York, 1962, p.
169).

\bibitem{4} L. D. Landau and E.M. Lifshitz, \textit{The Classical Theory of Fields}
(Pergamon Press, 1987, p. 280).

\bibitem{5} A. Papapetrou, Proc. R. Irish. Acad. \textbf{A52}, 11 (1948).

\bibitem{6} P. G. Bergmann and R. Thomson, Phys. Rev. \textbf{89}, 400 (1953).

\bibitem{7} C. M\o ller, Ann. Phys. (NY) \textbf{4}, 347 (1958).

\bibitem{8} S. Weinberg, \textit{Gravitation and Cosmology: Principles and
Applications of General Theory of Relativity} (John Wiley and Sons Inc., New
York, 1972, p. 165).

\bibitem{9} A. Qadir and M. Sharif, Phys. Lett. \textbf{A167}, 331 (1992).

\bibitem{10} C. M\o ller, Nucl. Phys. \textbf{57}, 330 (1964); K. Hayashi
and T. Shirafuji, Phys. Rev. \textbf{D19}, 3524 (1979); J.M. Nester, Lau Loi
So and T. Vargas, Phys. Rev. \textbf{D78}, 044035 (2008); Gamal G. L. Nashed
and T. Shirafuji, Int. J. Mod. Phys. \textbf{D16}, 65 (2007); J.W. Maluf,
F.F. Faria and S.C. Ulhoa, Class. Quantum Grav. \textbf{24}, 2743 (2007);
J.W. Maluf, M.V.O. Veiga and J.F. da Rocha-Neto, Gen. Rel. Grav. \textbf{39}%
, 227 (2007); A.A. Sousa, R.B. Pereira, A.C. Silva, Grav. Cosmol., \textbf{16%
}, 25 (2010); M. Sharif, Sumaira Taj, Astrophys. Space Sci. \textbf{325}, 75
(2010); Liu Y.X., Zhao Z.H., Yang J., and Duan Y.S., arXiv:0706.3245 [gr-qc].

\bibitem{11} K. S. Virbhadra, Phys. Rev. \textbf{D41}, 1086 (1990); K. S.
Virbhadra, Phys. Rev. \textbf{D42}, 2919 (1990); N. Rosen and K.S.
Virbhadra, Gen. Rel. Grav. \textbf{25}, 429 (1993); K. S.Virbhadra and J. C.
Parikh, Phys. Lett. \textbf{B331}, 302 (1994); J. M. Aguirregabiria, A.
Chamorro and K. S. Virbhadra, Gen. Rel. Grav. \textbf{28}, 1393 (1996); S.
S. Xulu, Mod. Phys. Lett. \textbf{A15},1511 (2000); S. S. Xulu, Int. J.
Theor. Phys. \textbf{39}, 1153 (2000); S. S. Xulu, Int. J. Theor .Phys.,
\textbf{46}, 2915 (2007); P.K. Sahoo, K.L. Mahanta, D. Goit, A.K. Sihna, S.
S. Xulu, U.R. Das, A. Prasad and R. Prasad, Chin. Phys. Lett. \textbf{32(2)}%
, 020402 (2015); S.K. Tripathy, B. Mishra, G.K. Pandey, A.K. Singh, T.
Kumar, S.S. Xulu, Adv. High \ Energy Phys. 705262 (2015)

\bibitem{12} I-Ching Yang and I. Radinschi, Chin. J. Phys., \textbf{42(1)},
40 (2004); I. Radinschi, Th. Grammenos, Int. J. Theor. Phys., \textbf{47(5)}%
, 1363 (2008); I-Ching Yang, Chi-Long Lin, I. Radinschi, Int. J. Theor.
Phys., \textbf{48(1)}, 248 (2009); I. Radinschi, F. Rahaman and A. Ghosh,
Int. J. Theor. Phys. \textbf{49}, 943 (2010); I. Radinschi, F. Rahaman, A.
Banerjee, Int. J. Theor. Phys., \textbf{50(9)}, 2906 (2011); I. Radinschi et al, Int.J.Theor.Phys. 51 (2012) 1425-1434 ;  M. Abdel-Megied, Ragab M. Gad, Adv. High Energy Phys.
\textbf{2010}, 379473 (2010); Ragab M. Gad, Astrophys.Space Sci. \textbf{346,%
} 553 (2013); T. Bringley, Mod. Phys. Lett. \textbf{A17}, 157 (2002); I. Radinschi et al, Int.J.Theor.Phys. 52 (2013) 96-104 ;  M.
Sukenik and J. Sima, arXiv:grqc/0101026; M. Sharif and Tasnim Fatima,
Astrophys. Space Sci. \textbf{302}, 217 (2006); M. Sharif, M. Azam, Int. J.
Mod. Phys. \textbf{A22}, 1935 (2007); P. Halpern, Astrophys. Space Sci.
\textbf{306}, 279 (2006); E. C. Vagenas, Int. J. Mod. Phys. \textbf{A18},
5781 (2003); E. C. Vagenas, Mod. Phys. Lett. \textbf{A19}, 213 (2004); E. C.
Vagenas, Int. J. Mod. Phys. \textbf{D14}, 573 (2005); E. C. Vagenas, Mod.
Phys. Lett. \textbf{A21}, 1947 (2006); T. Multamaki, A. Putaja, E. C.
Vagenas and I. Vilja, Class. Quantum Grav. \textbf{25}, 075017 (2008); L.
Balart, Mod. Phys. Lett. \textbf{A24}, 2777 (2009); Amir M. Abbassi, Saeed
Mirshekari, Amir H. Abbassi, Phys. Rev. \textbf{D78}, 064053 (2008); J.
Matyjasek, Mod. Phys. Lett. \textbf{A23(8)}, 591 (2008).

\bibitem{13} R. Penrose, Proc. R. Soc. London, \textbf{A381}, 53 (1982).

\bibitem{14} K.P. Tod, Proc. R. Soc. London, \textbf{A388}, 457 (1983)



\bibitem{14a} L.B. Szabados, Living Rev. Relativity \textbf{12}, 4 (2009) .
\bibitem{16} K. S. Virbhadra, Phys. Rev. \textbf{D60}, 104041 (1999).

\bibitem{17} Gamal G. L. Nashed, Mod. Phys. Lett. \textbf{A22}, 1047 (2007);
Gamal G. L. Nashed, Chin. Phys. Lett. \textbf{25}, 1202 (2008); Gamal G. L.
Nashed, Int. J. Mod. Phys. \textbf{A23}, 1903 (2008); Gamal G.L. Nashed,
Chin. Phys. \textbf{B19(2)}, 020401 (2010); Gamal G. L. Nashed, Int. J. Mod.
Phys. \textbf{A25}, 2883 (2010); M. Sharif, Abdul Jawad, Astrophys. Space
Sci., \textbf{331}, 257 (2011).

\bibitem{18} Chia-Chen Chang, J. M. Nester and Chiang-Mei Chen, Phys. Rev.
Lett.\textbf{\ 83}, 1897 (1999); C. M. Chen and J. M. Nester, Grav. \&
Cosmol., \textbf{6}, 257 (2000); Lau Loi So, J. M. Nester, Hsin Chen,
in \textit{Gravitation and Astrophysics: Proceedings of 7$^{\text{th}}$ Asia-Pacific Intern. Conf.}, (eds.) J.M. Nester, C-M. Chen, J-P. Hsu (World Scientific, 2007, p. 356); J. M.
Nester, Chiang-Mei Chen, Jian-Liang Liu and Gang Sun, in \textit{Relativity and Gravitation -- 100 years after Einstein in Prague}, (eds.) J. Bi\v{c}\'ak and T. Ledvinka (Springer, 2014, p. 177).




\bibitem{19} L. Balart, E. C. Vagenas, Phys. Rev. \textbf{D 90}, 124045
(2014).

\bibitem{20} H. Culetu, Int. J. Theor. Phys. \textbf{54(8)}, 2855 (2015)

\bibitem{21} E. Ay\'{o}n-Beato and A. Garc\'{\i}a, Phys. Lett. \textbf{B}
\textbf{464}, 25 (1999).


\bibitem{22}  I. Radinschi, F. Rahaman, Th. Grammenos, A. Spanou, and S. Islam, Adv. Math. Phys., \textbf{2015},   530281 (2015) e-Print: arXiv:1404.6410[gr-qc] .

\bibitem{23} J. M. Cohen, F. de Felice, J. Math. Phys. \textbf{25}, 4,
992 (1984).

\bibitem{24} K.S. Virbhadra and C.R. Keeton, Phys. Rev. \textbf{D77}, 124014
(2008).



\bibitem{25} K.S. Virbhadra, Phys. Rev.\textbf{\ D79}, 083004 (2009).




\end{thebibliography}
\end{document}